# Conscription and its exemption in 19th Century Japan: Incentivized family head in educational market.

YAMAMURA, Eiji

Department of economics, Seinan Gakuin University

e-mail: yamaei@seinan-gu.ac.jp

**Abstract**

Immediately after the establishment of the New Meiji Government in the 19th century, a system of conscription was adopted. The exemption rule has changed several times. Using individual-level panel data on the academic performance of Keio Gijuku, I found a surge in the family head's student rate between 1884 and 1888, and the rate declined immediately thereafter. After regaining privileges for private school students, family head performance declined, and the difference between head and non-family heads disappeared. This made it evident that conscription increased educational attendance quantitatively, but did not qualitatively improve academic performance.

# Introduction

Conscription is a type of market regulation that influences economic efficiency (Mulligan & Shleifer, 2005). Existing studies have explored how conscription influences educational attainment and outcomes in the labour market (Bound and Turner 2002; Hubers and Webbink 2015; Lemieux and Card 2001; Mouganie 2020; Torun and Tumen 2016). However, evidence on the relationship between conscription and educational outcomes is mixed. First, researchers have provided evidence that compulsory military service has a detrimental effect on educational attainment (Cipollone and Rosolia 2007; Hubers and Webbink 2015; Keller et al. 2010). Second, conscription improves educational outcomes (Angrist and Chen 2011; Card and Cardoso 2012; Card and Lemieux 2001; Maurin and Xenogiani 2007; Mouganie 2020; Savcic et al. 2023; Torun and Tumen 2016). A positive impact of conscription was found only among individuals with low cognitive skills (Bingley et al. 2022). Third, conscription had no effect on educational attainment (Di Pietro 2013) or earnings (Bauer et al. 2014; Grenet et al. 2011).

Young men are generally subject to conscription, which may affect the number of students in tertiary education. On the one hand, working in the military service reduced the time spent on educational learning. In other words, the relationship between conscription and education is simply a substitute. On the other hand, they were strongly motivated to avoid conscription, partly because the opportunity cost of conscription was very high. If students are allowed to defer conscription, eligible people will be motivated to enter school. Consequently, conscription increases the rate of school attendance[1].

However, these opposing views do not consider the following mechanism: After entering school, the students who aimed only to avoid conscription attained their goals. Therefore, they have very low motivation to learn and attend school to avoid dropping out. Conscription increased quantitative but not qualitative educational attendance However, most studies have not investigated this hypothesis. This study scrutinises this using a large sample of individual-level data in a novel historical setting as a natural experiment.

In studies analysing the mechanisms through which conscription affects human capital formation, researchers have taken advantage of various novel settings. Many studies have used lotteries as a randomisation tool to assess the causality between conscription and its educational outcome (Angrist and Chen 2011; Bingley et al. 2022; Siminski and Ville 2012). As a natural experiment, exogenous changes in the conscription rule (Savcic et al. 2023; Torun and Tumen 2016) or the abolishment of conscription (Card and Lemieux 2001; Maurin and Xenogiani 2007; Mouganie 2020) were used to identify causality. After the first adoption of conscription in 19th century Japan, the requirements for conscription exemption were revised several times within ten years. Considering this as an exogenous change in setting and a natural experiment, this study identified the effect of conscription exemption on educational attainment. In contrast to existing studies, the study deals with the period when conscription was first introduced in 19th century Japan. Due to frequent revisions of the conscription rule in the period 1884-1888, only the family head could be exempted from conscription if they were student at a private school because their privilege to be exempted was removed. This setting is considered a natural experiment that is useful for examining the influence of conscription on educational attainment.

---

[1] As another argument about positive effect of conscription, rather than draft avoidance, subsidies to veterans lead to educational gain (Angrist and Chen, 2011).

For the purpose of our analysis, we categorize the institutional changes into two distinct regimes: the 'Head Exemption period' (1884–1888), during which family heads were granted military immunity, and the 'No Exemption period' (1889–1898), following the abolition of this privilege. Standardizing these terms allows for a clearer comparison of student behavior across different incentive structures.

Directly after the Meiji Restoration in the 19th century, a conscription system was adopted in Jan, 1873, although conscription was abolished during the process of democratisation under occupation by the General Head Quarters (GHQ) after World War II (Gordon 2014, pp. 227-232). When the system was introduced, an exemption from conscription could be applied to those who met various requirements, for instance, students, family heads, the physically unfit, bureaucrats, criminals, and so on. Hence, in 1879, persons subject to conscription were about 320,000; however, 287,000 were exempted from military service (Oe 1981). Hence, exact enforcement strength was approximately 10%. The conscription rule was revised twice in the 1880s. In August, 1883, exemptions for private schools were eliminated. This has markedly reduced the number of private school applicants (Keio Gijuku 1958). However, the family head retained this privilege of being exempt from conscription. In February, 1889, the rule was further revised to abolish exemptions for family heads, and public and private school students were temporarily exempted from military service until the age of 26 (Ando 1979). Hence, private school students enjoyed the privilege of avoiding conscription.

According to Japan's family law, the status of the family head could be transferred to another family member if the current family head agreed. Furthermore, people could become family heads by being adopted into another family, such as by childless relatives or parental acquaintances. Hence, applicants in private schools had an incentive to become family heads if only family heads were exempt. However, abolishing exemptions for family heads reduced the incentives.

At Keio Gijuku School, known as the leading private school in Japan, all students' academic records were preserved, covering the period of rule change. Furthermore, the enrolment list was also reserved, which includes records of whether the students were family heads. By connecting these sources, we constructed an individual-level dataset that included over 6,500 individuals and their academic records in every term during their school attendance. Consequently, the sample size was approximately 40000. Based on the originally constructed large-sample historical dataset, using the difference-in-differences (DID) method and duration analysis (competing risk model), this study aims to explore the impact of rule change on exemption for conscription on the quantity and quality of students using the individual-level dataset of a private school. Previous studies have found heterogeneous effects of conscription on educational outcomes, which vary according to individual attributes (Bingley et al. 2022; Lemieux and Card 2001). Hence, I also scrutinised whether the effect of the rule changed according to whether students were Samurai and/or Tokyo residents, because their expected returns from education differed.

I found an increase in the number of students who were family heads from 1884 to 1888. Directly after the abolishment of family head exemptions, the rate clearly reduced. Compared to before the period of exemptions for family heads, the academic performance of students who were family heads improved, and their probability of dropping out declined.

The transaction cost of becoming the family head may be high because of the difficulty of coordination. Therefore, only promising young people can reduce the cost of reaching agreements within or between family members. Furthermore, they are strongly motivated to exhibit high academic performance. Overall, when private school students lost the right to avoid conscription, exemptions increased the family head's incentive to learn about their educational achievements.

Some existing studies have used historical data to quantitatively analyse the impact of conscription; however, after 20th century, society developed to be more systematic than before (Bound and Turner 2002; Cousley et al. 2017; Lindo and Stoecker 2014; Siminski 2013; Siminski and Ville 2011; Ville and Siminski 2011). Cross-country studies are useful for analysing conscription by comparing different institutional and economic conditions (Jehn and Selden' 2002; Poutvaara and Wagener 2007). However, these studies did not consider sufficient features such as law and family structure. The contribution of this study is to examine the relationship between conscription and its outcome, that is, educational attainment and family structure. I use large individual-level data from the era directly after entering the modernisation age in the far east non-Western country of Japan.

The remainder of this paper is organised as follows: Section 2 provides an overview of the situation during the study period. Section 3 describes the dataset used in this study. Section 4 proposes testable hypotheses and estimation strategies. Section 5 reports the estimation results and their interpretations. Section 6 discusses the findings of the study. In the final section, implications based on the findings are presented.

## Overview of Conscription in 19th Century Japan and Private Schools

In Japan, during the Meiji restoration in the mid-19th century, former Samurai people were deprived of not only economic privileges but also of wearing swords to deconstruct the feudal hierarchical structure of the Edo era. As part of the social system reform, the new Japanese Government first introduced a universal conscription system to recruit male adults of ages 20-40 for the imperial army. The conscription system was enacted in 1973 and has been in use since 1975.

<Insert Table 1>

However, as shown in Table 1, eligible individuals could avoid conscription if they met the exclusion criteria. When the system commenced, there were various cases of formal exemptions. For instance, students and teachers in both private and public schools, the eldest son, household heads, criminal persons, government officials, and the physically unfit were exempt from conscription. Furthermore, people could buy right to be exempted if they were compensated with a huge fee of 270 yen, which is higher than the annual earnings of ordinary people (Gordon 2014, pp. 66-67). Furthermore, the new government planned to identify eligible persons based on the records in the family register ('Koseki daicho'). In reality, the record was incomplete, and the system was loose. In addition, until the end of the 1870s, internal wars occurred frequently between a part of the former Samurai who strongly opposed conscription and the new government, because they had lost their privilege of wearing their sward[2]. Under chaotic and unstable circumstances, avoiding conscription was easy (Ichikawa 2022). Only 3% of eligible persons were actually conscripted in 1875 (Fujiwara 1975).

In the first period that was defined, the critical point was that the Keio students could avoid conscription unless they dropped out. However, the rule changed in December 1883; entering the second period, privileges for private school students were abolished; thus, Keio students were not exempted. This change caused the number of private school students to decrease substantially (Sekiguchi, 2008).

<Insert Figure 1 >

This caused private schools such as Keio Gijuku to be in predicament to close school (Keio Gijuku 1958). Consistent with this, Figure 1 shows that the number of Keio Gijuku students declined from 1883 to 1886. However, the negative shock of the rule change did not persist, and students increased distinctly again from 1887 onwards.

Even during this period, the family head was exempted. Therefore, many eligible persons strategically became family heads, who were legally and formally approved (Ichikawa 2022).

---

[2] Wars had occurred especially in Southern part of Japan, in Kyushu region. For instance, Saga war (1874), Akizuki war (1876), Keishinto war (1876), and Seinan-war (1877). Seinan-war was the largest rebellion against the Maiji government directed by Takamori Saigo who belonged to the Samurai class in the Edo-era and one of the three great nobles who led the Meiji Restoration.

Various books on how to be exempt from conscription have been published (Kato 2000). Many Keio students were able to manipulate their personal records to enjoy exemptions. For instance, they changed the name and record of a family register to meet the exemption requirement. Usually, the eldest son of the family becomes the family head if the former head, ordinarily the father, passes away. However, if the former head agreed to transfer his family status, the son could become the family head even when his father was alive. To become the family head, the non-eldest son was adopted by the family of a childless relative. His name was changed if the relative's name was different. In the dataset used in this study, those who changed names were 4.7% for family heads and 3.5% for others. This indicates that changing the name was used to avoid conscription. In addition to becoming the family head, there were various ways to avoid conscription (Ichikawa 2022). Changing the name was acknowledged as a technique to avoid conscription to another ineligible person or escape detection by military offices.

In January 1889, after a further revision of the Conscription Ordinance, the exemption for family heads was abolished (Ando 1979). In other words, the government modified the rules to close loopholes. Fortunately, however, for Keio Students, private school students could regain the privilege of being temporarily exempt from military service until 26 years of age (Ando 1979). Therefore, substantially only in a very short period (1884-1888), Keio students lost their privilege to be exempt from conscription, leading them to be incentivised to avoid conscription.

In the internal rule of Keio during the study period, students would automatically be treated as dropping out only if they did not attend school for over three months or if educational fees had not been compensated (Keio Gijuku Fukuzawa Research Center 1989a). In other words, academic performance and cognitive skills did not influence the probability of dropping out. Further, to foster the business elite, Keio provided and focused on practical science, such as accounting—exported from Western countries[3]. Academic performance, such as ordered rank in class, reflects the degree of motivation to learn and increases students' returns from education.

Prior to reform of Meiji restoration, Samurai people received their salaries, called 'stipends', from their feudal lord 'Daimyo' (Gordon 2014, p. 17). As part of the reform, the new government announced a tax Samurai concerning their stipends at the time of the introduction of the conscription system. Furthermore, in 1876, it became compulsory for all stipends to be converted to interest-bearing, with a face value of an average of 10 years of income. Inevitably, most Samurai suffered substantial economic loss. However, they did not have a family business to encounter predicaments. The situation of Samurai was described through the following proverb: 'Sihzoku no Shoho', meaning, a samurai's way of business did not meet the requirement of society and so ended in failure (Uchiyama 2020).

---

[3] Actually, many graduates from Keio became business leaders in predominant firms in modern Japan (Keio Gijuku 1958).

## Data and Methods

Sources and constructing datasets

In a series of Fukuzawa-related historical documents, various individual-level datasets have been compiled for the Meiji period. In this study, I integrated two datasets to explore how the revision of the conscription rule influenced educational outcomes.

First, the academic record list, 'Gakugyo Kinda-Hyo', includes all students' academic records, such exam score, about various subjects and ordinal ranks in class every term over the period 1871-1898 (Keio Gijuku Fukuzawa Research Center 1989c). During this period, school attendance was divided into three terms in a year. Grades and subjects changed frequently because the educational system was modified through trial and error. It was difficult to compare the performances of the subjects. We obtained the number of students in a class and the ordinal rank in a class throughout the period. Recent studies have indicated that ordinal ranking is valuable for measuring students' academic attainment (e.g., Denning et al. 2023; Elsner et al. 2021; Elsner and Isphording 2017; Isozumi et al. 2021; Murphy and Weinhardt 2020; Pagani et al. 2021). Subsequently, the ordinal rank was used as a proxy for academic performance. Furthermore, the duration of attending Keio was also identified from the data.

Second, all applicants were obliged to fill in the blank spaces for admission to Keio School. The documents were compiled to be lists of freshmen-students, 'Nyu-sha cho', covering from 1863 to 1901 (year of the death of 'Yukichi Fukuzawa' Founder of Keio). I collected information regarding the name and his social status 'Samurai or Ordinary citizen', and address of hometown from the list (Keio Gijuku Fukuzawa Research Center 1989b). The list contained information on individuals formally registered as Keio students, even if they dropped out immediately after entering Keio.

As explained earlier, initially, the conscription system did not function, and society was unstable because of internal wars. In addition, various factors influenced the number and performance of students. Therefore, we focused on peaceful and stable times as long as the dataset was covered. Thus, we investigated the period of 1880-1898 in this study.

The constructed dataset includes 9,731 individuals and 49,463 observations. Furthermore, the students belonged to 1,672 classes. The class is considered a unit for determining the ordered rank, which was a measure of academic attainment in this study. However, some basic variables were unavailable. Documents and records from the Meiji era were written in hand. In some cases, the style was too different from the present-day style to decipher scripts. Furthermore, some students were not assigned to a class if they were irregular. These observations were excluded from the sample. The sample size of the dataset used for the estimation was reduced to 6,700 individuals and 3,9694 observations although the number of classes was maintained at 1,672.

Measurements

< Insert Table 1 >

In Table 2, I describe the variables used in the estimation and their basic statistics, such as mean values and standard errors. In the data, mean value of '*Class size*' was measured by the number of students in the class where students belonged. The ordinal rank in a class depended on the class size, because the bottom rank is second in the two-student class, while it is $50^{th}$ in a class with 50 students. Therefore, to control number of students in a class, the rank order should be standardised as 'Ordinal rank/Class size'. Hence, the bottom rank is one in this definition, regardless of the class size. Further, for convenience of interpretation, '1- Ordinal rank/Class size' is used as *Academic Performance*. The values were in the range of 0 – 1. The larger the value, the better the performance.

Skipping was included as an alternative proxy for educational attainment, *Skip* is included. In Keio, students skipped grades to receive higher grades if their academic achievement was excellent. I obtained this information from the records. The dummy variable for *Skip* was used as a robustness check.

During the period 1884-1888, students were manipulated to become family heads to avoid conscription, which was captured by *Head*. In addition, changing one's name was used to be exempted in different ways (Ichikawa 2020). I identified students who changed their names using corresponding lists of freshmen and academic records. However, the timing of the name change was ambiguous. Therefore, different from *Head* which could be identified at the entrance to Keio, whether changing names was influenced by rule changes remained ambiguous. However, *Name Change* could be used as a control variable.

Preliminary analysis

<Insert Figure 2 >

Figure 2 shows the changes in the percentage of family head freshmen students among all freshmen. In the first 'loose' period, family head rate was very low, around only 1 %, reflecting that there was no need to become a family head to be exempted. That is, young males unintentionally became family heads even though they were not mature enough, mainly because unexpected events occurred, such as the father's early death. Surprisingly, in second 'head' exemption period, directly after deprivation of Keio student's privilege to avoid conscription, family head rate increased drastically and constantly (to around 13%)—almost 13 times larger than before rule change. This reveals that applicants had a strong motivation to become family heads to be exempted.

After the Keio student regained privileges in the no-exemption period, the family head rate declined. However, there is time lag to decline while the rate did not turn back to 1%. The transaction cost of strategically becoming the family head was considered very large. In most cases, the former family head might be alive, and other family members and relatives might expect him to dominate as the head. Replacing the family head necessitated not only the applicant's intention, but also various persons' agreements. For this purpose, the applicant must persuade the current family head and family members to accept him as the new head. Inevitably, prior to applying for school, it took several years to become a family head. Several years before unexpected rule change, a part of applicants might have been family heads, so the family head rate stabilised around 7%.

<Insert Figures 3 and 4 here>

Figures 3 and 4 compare the academic performance of family and non-family heads in different periods. The time point of rule changes occurred in December 1883 and January 1889. Accordingly, applicants were strongly incentivised to become family heads in the period 1884-1888. Thus, the group can be divided into three periods 1880-1883, 1884-1888, and 1889-1898, represented by Figures 3-5, respectively, and Tables 3-6, A1, A2 presenting the regression estimation results.

From Figures 3 and 4, in the period of 1880-1883, the academic performance of the family head was lower than that of the non-head, despite not being statistically significant. In particular, nobody could skip for group of family heads. Wide range of 95 % confidence interval implied a larger standard deviation, that reflects that a few students were family heads (Figure 2).

In the period 1884-1888, the family head's academic performance outweighed that of the non-family head, and the difference was statistically significant. Subsequently, in the period 1889-1898, the difference in performance disappeared. The family head's incentive to learn declined to the non-head level.

<Insert Figure 5>

Figure 5 compares the dropout rates between groups and periods. In the period 1880-1883, dropout-rate of family heads was 40%, which was far higher than the 15% for non-head students. However, in the period 1884-1888, the rate declined drastically to approximately 10%, and difference between groups disappeared. Considering Figures 3-5, the non-family heads attended school to avoid dropping out, but were less likely to make efforts to learn than the family head. In the period 1889-1898, the difference between them disappeared, like in Figures 3 and 4.

## Hypotheses and estimation strategy.

Hypotheses

The deprivation of private school students' privileges gave applicants a strong incentive to be exempt from conscription. Naturally, the students strategically manipulated this to become the family head. However, for this purpose, the current family head (at the moment) and other members must accept the proposal. For these individuals, the expected return from replacing the family head was sufficiently high. Therefore, a young boy was required to exhibit competence and promise to meet this requirement. Even after being approved as the head and entering Keio, he was greatly motivated to learn and master practical science to increase return from education. Furthermore, based on Figures 3-4, I propose *Hypothesis 1*.

*Hypothesis 1: Family head students' academic performance was better than other students if they entered Keio in the period in which the privilege to be exempted was removed for students of private schools.*

Keio students were less likely to endeavour to improve their skills if private school students had the privilege of being exempted. Participants were forced to drop out if they had been absent for several months. Low academic performance prevented them from achieving higher grades at best. However, to maintain their privilege, they were incentivised only to attend school, which did not substantially improve their performance. If they could avoid conscription for reasons unrelated to school, their motivation to avoid dropping out decreased. In other words, Keio students' incentives to attend school became stronger if they could not avoid conscription for other reasons. Accordingly, I propose *Hypothesis 2*.

*Hypothesis 2: Keio students were less likely to drop out if the status of being a school students enabled them to be exempted from conscription.*

During the Meiji era, various economic resources were brought to Tokyo. New firms, such as mushrooms, appeared in Tokyo. The benefits of an agglomerated economy increased business opportunities in the city. Therefore, the expected returns from learning practical science were greater for students born in Tokyo, where their family businesses were located. To compensate for the educational fee, Keio Students must be born into wealthy families, even if they were commoners. During the Meiji era, the family's wealth came from their family business, and Keio students were expected to develop the family business to fit modernisation. The family head had the right to conserve and dispose of the family's assets and manage businesses. The expected return from education was higher for commoner students than Samurais if students' hometowns were Tokyo than if their hometowns were not Tokyo.

Hence, I propose *Hypothesis 3*:

*Hypothesis 3: Family heads of commers born in Tokyo were more likely to exhibit better performance than Samurais.*

Estimation Strategy

The dataset covers the period of 1871-1898. The period was divided into three parts according to the change in the rule of exemption (See Table 1). As discussed in a previous section, Figures 1-5 indicate that the three periods corresponded with different phases. To examine the effect of the rule change, the interaction terms of *Head* and period dummies (*Head Exemption*, and *No Exemption*) were included in the full sample estimations in the OLS and Duration analyses explained in the following subsection.

Estimation about ordered rank in class

The baseline estimated model is as follows:

$$Academic\ Performance_{it}$$

$$= \alpha_0 + \alpha_1 Head_i \times Head\ Exemption_t + \alpha_2 Head_i \times No\ Exemption_t$$

$$+ \alpha_3 Head\ Exemption_t + \alpha_4 No\ Exemption_t + \alpha_5 Head_i$$

$$+ \alpha_6 Name\ Change_i + e_s + \varepsilon_{it}$$

In this specification, *Academic Performance $_i$* represents the dependent variable for individual $i$ and its value ranges from one to zero. $\alpha$ denotes the regression parameters. In the alternative specification, *Skip* was included as the dependent variable. $e_s$ is the fixed effect for class $s$ to which the individual belonged to. Error term is $\varepsilon_{its}$ in time point (academic term) $t^4$. Simple Ordinary Least Square (OLS) was used with 1,677 class dummies using 39,694 observations. As shown in Table 2, the dependent variable had an upper limit of 1. In this case, the Tobit model was preferred. Nonetheless, the number of class dummies was too large to obtain results because the calculation could not converge in the maximum-likelihood method.

*Head* was the information at Keio and so did not change during the period when he attended school. Estimates of the *Head* could not be calculated if the Fixed Effects (FE) model was used to control for the students' time-invariant characteristics. Instead, we added 1672 dummies to control the feature of class to further control for grade, set of subjects, and peer effect. That is, the model was considered a fixed-effects model to control the characteristics of the class.

To test *Hypothesis 1,* I examined the coefficients of $Head_i \times Head\ Exemption_t$. The predicted sign of $\alpha_1$ was positive. To test *Hypothesis 3,* I added triple-cross terms to the baseline model. The key variables were $Head \times Head\ Exemption \times Tokyo$ and $Head \times Head\ Exemption \times Commoner$. Based on *Hypothesis 3,* the coefficients of these variables were expected to be positive.

The type of specification was DID. A parallel trend condition was required to show that the DID was valid for the examination. In the appendix, Figures A1 indicates the trend in *Academic*

---

[4] Apart from variables in the model, family head rate in a class is included but any of results show statistical significance.

*Performance* for family and non-family headed students, covering the period before and after the rule change. Owing to the very small sample size, the slope of the family head's performance was steeper than that of non-heads before the revision of the rule. However, both groups showed similar tendencies, first increasing and then slightly until 1883. Therefore, the DID method was valid to a certain degree.

Competing-risk regression for duration of attending school.

In addition to the ordered rank, I explored the probability of dropping out and how students survived attending school. As shown in *Hypotheses 2*, students might only have attended school rather than learn useful practical science if they aimed to evade conscription by entering school. After controlling for ordered rank, we scrutinised it to test *Hypothesis 2*. The estimated function was expressed as follows:

$$Pro(Drop\ out)_i = \beta_0 + \beta_1 Academic\ Performance_i + \beta_2 Head_i \times Head\ Exemption_t$$
$$+ \beta_3 Head_i \times No\ Exemption_t + \beta_4 Head\ Exemption_t + \beta_5 No\ Exemption_t$$
$$+ \beta_6 Head_i + \beta_7 Name\ Change_i + \beta_8 Class\ Size_i + u_i$$

A proportional hazards model has been employed in a previous study to examine dropout decisions (Li 2007). Considering the real situation, two different outcomes were considered as dropouts if the model was applied. In addition to dropouts, graduation also led to students leaving school. Thus, it was necessary to identify whether leaving school could be considered within the dropout rate. To solve this problem, a competing risk regression was preferable. In this model, dropping out is a failure of our primary interest, whereas graduation is treated as a competing failure. Therefore, I could investigate the probability of dropouts occurring after controlling for competing failures.

A record of academic performance was not compiled for a certain period in 1898. Even after this, many students were thought to have attended school, although there were no academic records. Identifying whether to drop out or continue attending school was impossible if students' records appeared in the last term of the list. Therefore, as a robustness check to deal with the bias, we conducted the estimation simply using a subsample where observations in 1898 were excluded.

As shown in the specifications, the effect of *Academic Performance* was controlled. The other independent variables were almost identical to those in the OLS model. One difference is that group dummies are not included, because the maximum-likelihood method could not converge if many dummy variables were included. Instead, *Class Size* was included to control for the effect of the class to which the students belonged.

We should pay attention to this difference from OLS model. In duration analyses, such as the competing-risk model, the hazard ratio (HR) was usually reported rather than the coefficient. The interpretation of the HR was different from that of the OLS coefficient. The HR value was always positive. However, it was important to determine whether the HR value was larger or smaller than one. If the HR value was greater than one, the variable increased the probability of dropping out. If the HR value was less than one, the variables reduce the probability of dropping out.

From *Hypothesis 1*, the HR of *Head×Head Exemption* was anticipated to be smaller than one. Based on *Hypothesis 2, No Exemption* was expected to reduce the dropout probability. Therefore, the HR of the linear combination of *No Exemption* and *Head×No Exemption* was expected to be smaller than one. Similar to the OLS model, alternative specifications and triple cross terms with *Tokyo* or *Commoner* were added. From *Hypothesis 3*, the predicted HR of these triple-cross terms are smaller than one.

**Estimation results and its interpretation**

OLS estimation regarding educational attainment

<Insert Table 3>

Table 3 presents the estimation results for the baseline model. The dependent variables are *Academic Performance* and *Skip* in columns (1) and (2), respectively. The key variable *Head × Head Exemption,* produced the expected positive sign while being statistically significant. This indicates that the academic attainment of family head students during the period when the family head was required to avoid conscription was better than when all private school (Keio) students were allowed to do so. Value of the coefficient is 0.118 in column (1), meaning that family head's ordered rank in a class was 11.8 % higher than family head in the former period. In Column (2), the value was 0.037, implying that family head was more likely to skip a grade by 3.7%. The effect of rule changes on the academic outcomes of the head was sizable. This finding is consistent with *Hypothesis 1. Head ×No Head Exemption* also shows a significant positive sign, meaning that the positive effect of family heads persisted even after the Keio students regained privilege. However, the values of its coefficient are smaller than *Head ×Head Exemption*. Hence, the effect of family heads persisted, although the effect was reduced. Some students were approved as family heads before the rule was revised. Their motivation to improve their skills was thought to be high; therefore, the effect of the family head did not disappear. A significant positive sign for *Name Change* implies that those who changed names exhibited better academic performance than other students. In a previous section, I introduced anecdotal evidence that students changed their names to avoid conscription. Similar to the case of family head students, they had a strong motivation to avoid conscription and learn, although the timing of changing names was unknown.

<Insert Tables 4 and 5 > here >

Tables 4-5 present results with triple cross terms. Here we put focus on key variables to test

*Hypothesis 3.* In Table 4, we observed the significant positive sign of *Head ×Head Exemption × Tokyo* in Column (1), despite being statistically insignificant in Column (2). This means that Family head students who were born in Tokyo exhibited higher ordered rank than those born in other places in the period when family head could avoid conscription.

In Table 5, a significant positive sign for *Head ×Head Exemption ×Commoner* was observed in Column (2), although it is statistically insignificant in Column (1). The probability of skipping a class for the family head of the commoner was higher than that for the Samurai. Considering the results in Tables 4 and 5, *Hypothesis 3* was supported.

Duration analysis regarding probability of dropout

<Insert Table 6>

Tables 6, A1 and A2 reported the results of duration analysis using the competing risk model. In Columns (1), (2), (4), and (5), *Academic Performance* or *Skip* were included to control for the academic achievement effect. To mitigate the problem of abolishing the academic list, I used a subsample excluding the last year of 1898, and the results are shown in Columns (1)–(3). The results obtained using the full sample were shown in Columns (4) to (6).

Table 6 presents the baseline model results. HR for *Academic Performance* and *Skip* were far smaller than one and statistically significant. Reasonably, the better the academic performance, the lower is the dropout rate.

As expected, HR of *Head ×Head Exemption* was approximately 0.4, which is smaller than one while being statistically significant at the 1% level. Family heads' probability of dropping out in the period when they could to be exempted was approximately 60% lower than that in the former period. This finding is consistent with *Hypothesis 1*. To test *Hypothesis 2*, I checked how students' dropout rates differed from those of others in the period after they regained the privilege of exemption. Figure 6 suggests that Keio students were less likely to drop out, even after regaining the privilege, than when they could avoid conscription through the status of being a student and other measures. As shown in Columns (3) and (6), the HR is smaller when the academic achievement effect was not controlled for than when it was controlled for in Columns (1), (2), (4), and (5). Owing to the inclusion of the proxy variable of academic achievement, the interaction effect decreased but did not disappear. That is, the probability of dropping out depended not only on academic achievement but also on other factors, as proposed in *Hypothesis 2*. These results are consistent with *Hypothesis 2.*

<Insert Figure 7>

Using the results of *Head ×Head Exemption* and *Head Exemption* in Column (4), I visually compared changes in dropout probability between family head students and other students in the period 1884-1888. Figure 7 illustrates the dropout incidence curves for the two groups. A cursory examination of Figure 7 reveals that the family head's probability of dropping out was clearly lower than that of non-family heads in the period.

To test *Hypothesis 3,* I now turn to Tables A1 and A2, which include the triple cross terms. Table A1 indicates that the HR of *Head ×Head Exemption ×Tokyo* ranged from 0.056 to 0.084, which is far smaller than one. Furthermore, it is statistically significant at the 1% level in all columns. This implies that the family head born in Tokyo has a far lower probability of dropping out than the head born in other places during the period in which the family head was required to be exempted.

In Figure A3, the dropout incident curves shows that heads born in Tokyo had a remarkably lower dropout probability than heads born in other places in the period.

In Table A2, we observed statistical significance for *Head ×Head Exemption ×Commoner* and its HR ranged between 0.189 and 0.313, which was smaller than one. From the results, in comparison with the head of the Samurai family, I argue that the head of the commoner family was less likely to drop out during a period in which the family head could be exempted.

Figure A4 demonstrates the dropout incident curves, indicating that the head of the common family showed a clearly lower probability of dropping out than that of the Samrai family in the period when the family head was required to avoid conscription.

Hence, as shown in Tables A1 and 2, Figures A3 and A4 strongly support the *Hypothesis 3*.

## Discussion

Before the revision of 1883, students had an incentive to learn in school to avoid conscription, leading to an increase in college attendance and degrees (Card and Lemieux 2001). However, the cost of entering Keio school was low for wealthy people. In fact, in the 19th Century, applicants were limited to wealthy people who could enter Keio without being examined. Therefore, those who only aimed to enjoy the privilege of exemption from conscription were less likely to be motivated to learn to improve their skills and competence.

Once private school privileges were removed, less motivated students disappeared. Meanwhile, promising young men could become family heads to avoid conscription because they were able and promising enough to be supported by family members and relatives. They entered school to learn not only for their own interests but also to meet the expectations of these supporters. Naturally, students with young family heads would have greater motivation than other students.

There are possibly two mechanisms that family heads are more motivated (an incentive effect) and that only promising individuals could become family heads (a selection effect). Since heads were chosen based on their potential as family leaders rather than their academic test-taking skills, the observed changes in their academic performance following rule changes are more likely to reflect shifts in their effort (incentives) rather than a pre-existing academic superiority. Furthermore, during this period when modern formal education had not yet become widespread, the family members responsible for designating the head typically lacked experience with modern schooling themselves. Consequently, we assume that it was practically impossible for them to discern or predict how a young man would be evaluated under a modern academic grading system after enrollment. On the assumption, we argue that an incentive effect is larger than a selection effect as plausible and dominant interpretation. The results, however, should be interpreted with caution due to potential selection on unobservables that we cannot fully control for with the data used for this study.

Regarding the concerns about parallel trends and small sample sizes in the pre-period, we have conducted an event-study analysis. The results are presented in Figure A2 (Appendix). We acknowledge that the number of family-head students in the pre-period (1880–1882) is extremely limited, totaling only 13 observations. Due to this data constraint, we used the entire pre-period as the reference (baseline) to stabilize the estimates. The small sample size increases the noise and widens confidence intervals for the early years, it does not invalidate the subsequent shifts observed.

As shown in Figure A2, the coefficients for the family-head status are positive across almost all years. Crucially, the coefficients become statistically significant primarily during the "Head Exemption period" (1884–1888), showing a higher frequency of significance compared to the "No exemption period." This pattern indicates that the performance advantage of family heads was most pronounced precisely when the institutional incentives were strongest.

One limitation of using normalized ordinal ranks is the potential for selective attrition. As students with poor academic performance are more likely to drop out, the remaining student body in upper grades becomes increasingly composed of higher-ability individuals. This process creates a more competitive environment, making it inherently more difficult for students to improve their relative rank compared to their time in lower grades, regardless of class size. Because this dynamic selection bias is not fully captured by our model, the rank-based results should be interpreted with caution and may represent a conservative estimate of the true incentive effect. Therefore, academic survival and skip outcomes are more absolute measures of achievement because these are less likely to be sensitive to the mechanical fluctuations of relative class ranking.

During the study period, the Samurai class lost economic privileges and wealth despite maintaining their social status. Wealthy commoners were required to modernise their family businesses. Hence, they were more motivated to learn the practical science provided by the Keio School than Samurai students. In Tokyo, the expected return from learning science was very large because modern firms were concentrated in the city as a consequence of the Meiji Restoration. Therefore, students who were raised in Tokyo had greater motivation to learn. Consequently, commoners brought up in Tokyo tended to display higher academic performance than others. This is an unintended outcome of a change in the conscription exemption rule.

Previous studies found that conscription had no influence on earnings (Bauer et al. 2012; Grenet et al. 2011). Evidence suggest that conscription increased educational attendance but did not improve wages (Mouganie 2020). Mouganie (2020) argued that the average marginal return on additional schooling induced by conscription is very low. Considering our estimations, conscription increased quantitative school attendance or graduation, but did not substantially improve the quality of their skills and competence, with the exception of family heads. Therefore, the human capital accumulated through conscription was limited to some groups under specific conditions. Overall, conscription was considered to be a less efficient outcome. The findings of this study reveal the underlying mechanisms.

## Concluding remarks

In Japan, the Meiji restoration in the mid-19$^{th}$ century, as part of the reform of the social system to deconstruct the hierarchical structure, was first adopted by deriving Samurai's privilege of wearing a sword. In the early period, the requirement to be exempt from conscription was revised several times in the 1880s, which was politically stable and peacetime after the suppression of the former Samurai riots and internal wars.

In a novel historical setting, this study investigated the following unexplored mechanism through which conscription influenced educational outcomes. Eligible individuals were motivated to attend school to avoid conscription. However, this does not mean they were motivated to learn once they enter school. The elite status and legal flexibility of the Keio Gijuku students provide a unique "laboratory" to observe how individuals strategically respond to institutional incentives. By documenting how students prioritized enrollment (to maintain exemption) without a corresponding increase in learning quality, we reveal a fundamental mechanism of "strategic school-staying" that can occur in any setting where academic credentials or enrollment status provide non-academic benefits (e.g., visa status, tax exemptions, or conscription avoidance).

We found that the family head students of private schools exhibited a higher ordered rank in class and lowered their probability of dropping out after restricting the exemption for family heads. However, the rule again changed to entitle the privilege of exemption to private school students, which led to a decline in the family heads' academic performance. Meanwhile, their dropout rate continued to be lower than before restricting the exemption for family heads, although there was no difference in dropout between heads and no-heads.

20th-century lottery-based studies explore the impact of actual service on human capital (e.g. Angrist and Chen, 2011; Bingley et al., 2022; Siminski, 2013; Siminski and Ville, 2011). Different from it, our study uncovers the distortionary effects of exemption rules on educational investment.

This distinction highlights a different dimension of the relationship between military institutions and education—one focused on the strategic behavioral responses to institutional loopholes.

Overall, the findings revealed that conscription did not substantially improve students' skills and competence, despite seemingly accumulating human capital measured by additional schooling years. The findings of this study are consistent with the argument that conscription induced longer school attendance, whereas the average marginal return on additional schooling was very low (Mouganie 2020). In conclusion, from the perspective of economics, it was far more critical for students to be motivated to improve their productivity than to obtain an academic background for nonproductive purposes. Therefore, rules should be designed to mitigate the issue.

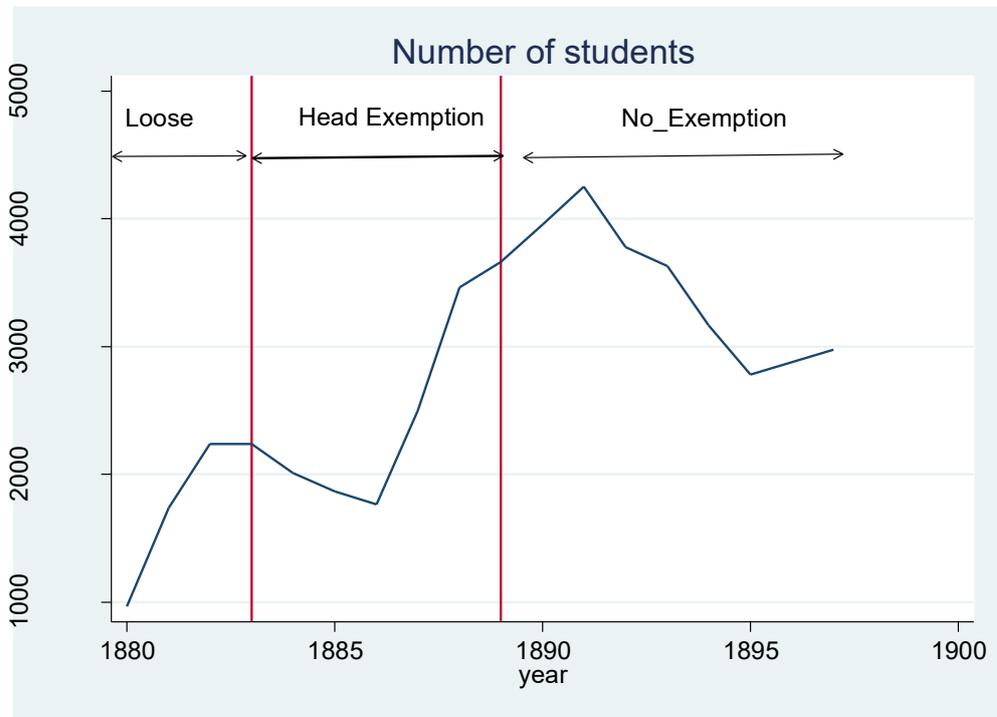

**Fig1. Change of Total Number of Students 1880-1898**

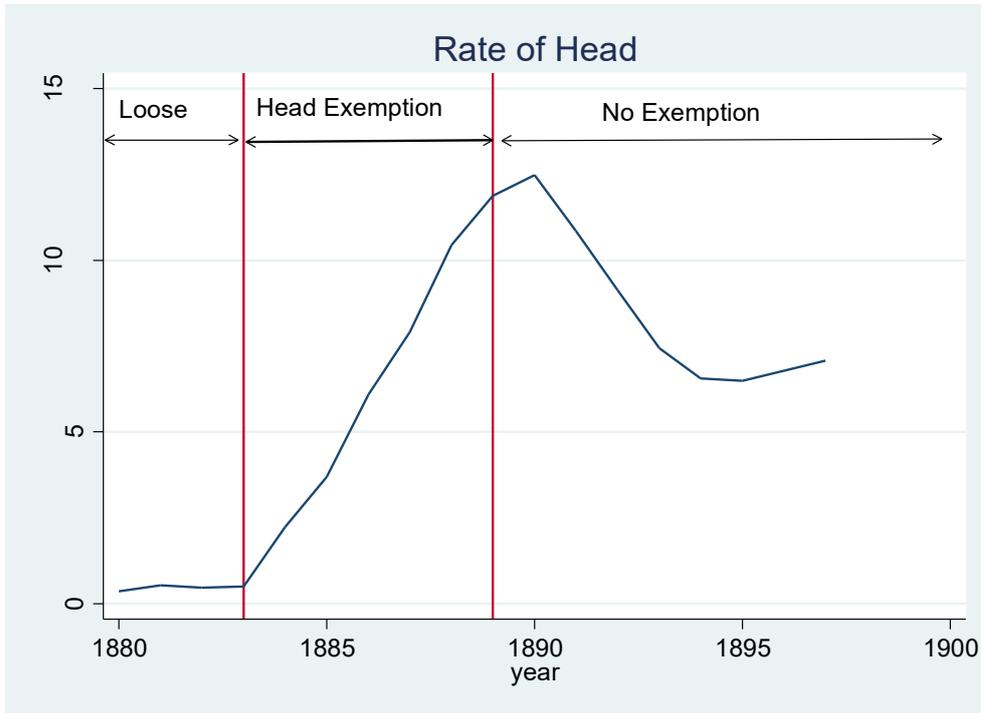

**Fig.2   Change of Head Students rate in Freshmen students in each year**

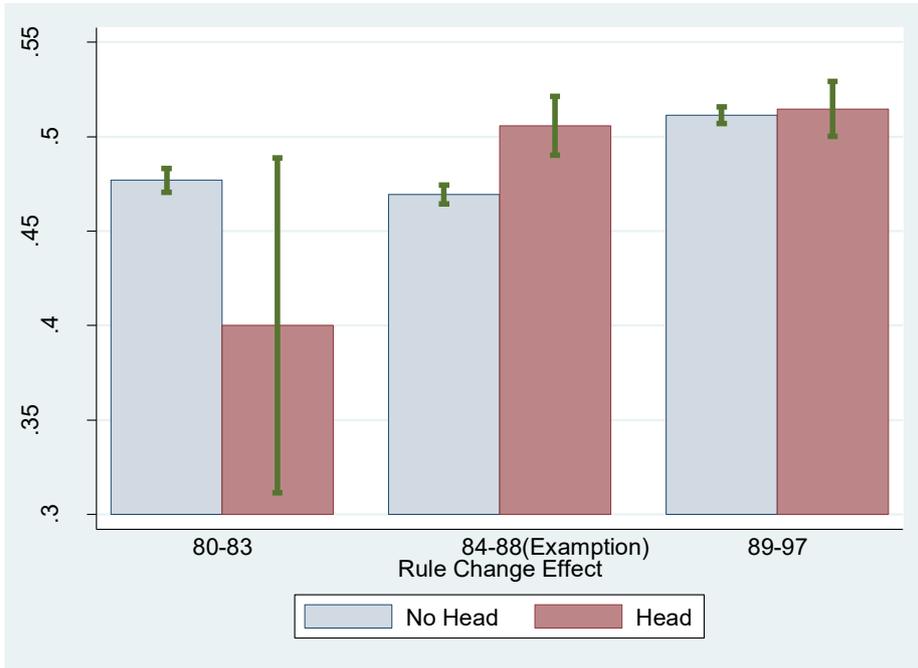

**Fig.3.** Academic performance: Head vs No-head.   Note: Group separated by years of entrance
Note: Group separated by years of entrance. Bars show 95% level confidence intervals.

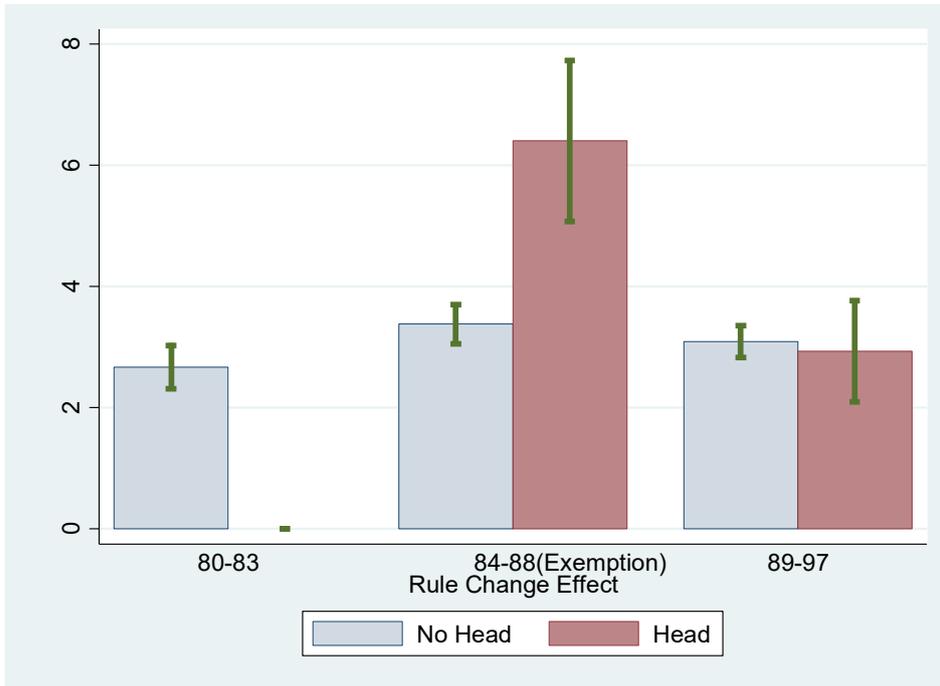

**Fig.4.** Skipping rate: Head vs No-head. Note: Group separated by years of entrance

Note: Group separated by years of entrance. Bars show 95% level confidence intervals.

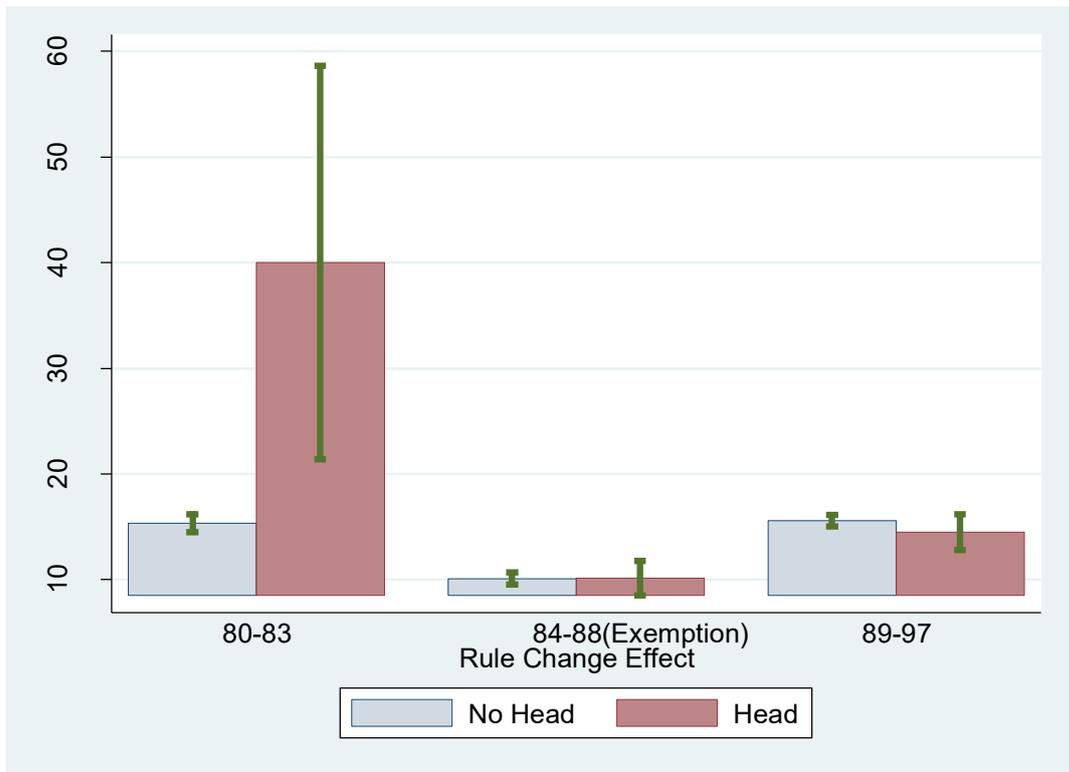

**Fig.5.  Percentage of dropout: Head vs No-Head**

Note: Group separated by years of entrance. Bars show 95% level confidence intervals.

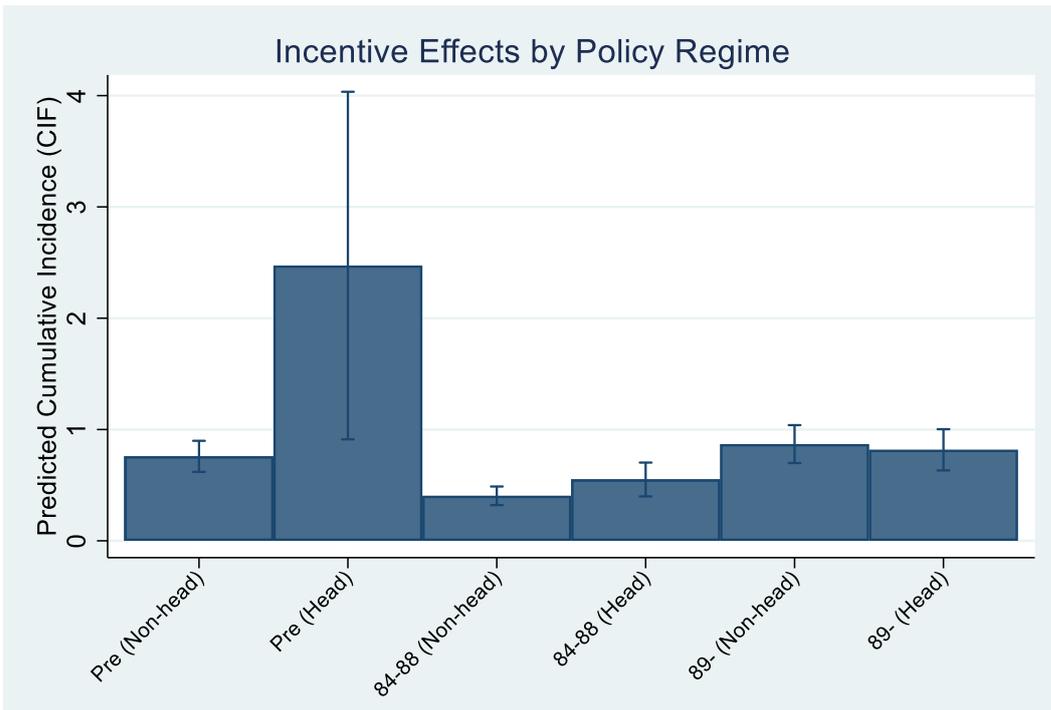

**Figure 6. Marginal effects about duration analysis of column (1) of Table 6.**

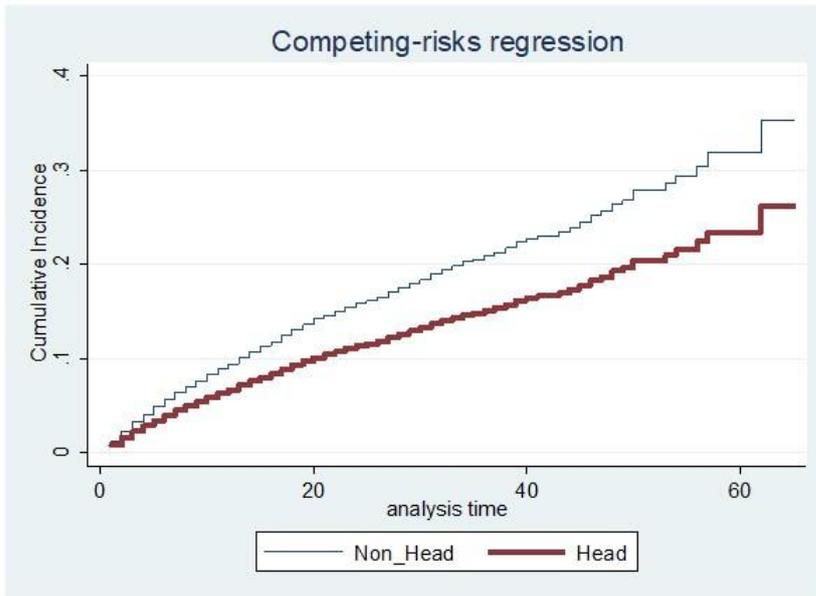

**Fig7. Drop-out Incidence Curve: Head vs No-Head 1884-1888 period**

(Drop-out and Graduation are as competing risks)

Note: Figure illustrated based on results of *Head×Head Exemption* and *Head Exemption* and column (4) in Table 6.

**Table 1.** Rule changes about exemption

|  | 1873-1883 | 1884-1888 | 1889- |
|---|---|---|---|
| Private school Student | ○ | × | ○ |
| Family Head | ○ | ○ | × |
| Other conditions where private school students could be exempted. | ○ | △ | × |

Notes: Female ages of 20-40 are a person on a conscription list. However, unfit persons could be exempted in terms of physical condition. Further, they could avoid conscription if they belonged to group of "○" in the Table. In the group of △, whether they could be exempted depends on the case. In group of "×", they could not be exempted.

**Table 2.** Description of variables and its mean value and standard deviation.

| | Description | Mean | S.d. |
|---|---|---|---|
| *Academic Performance* | Mean values standardized individual's ranking within the class during academic terms. Calculating "Rank" by "individual's ranking / number of classmates" in each academic term. For convenience of interpretation, the values is changed as "1−Ranking". | 0.489 | 0.290 |
| *Skip* | It is 1 if student skip a grade, otherwise 0. | 0.032 | 0.174 |
| *Head* | It is 1 if student was head of family, otherwise 0. | 0.074 | 0.262 |
| *Loose* | It is 1 if student entered the school before 1882 when conscription rule was loose with various kinds of exemptions, otherwise, otherwise 0. (default period in regression estimations) | 0.126 | 0.332 |
| *Head_ Exemption* | It is 1 if student entered the school between 1884-1888, head exemption period, otherwise, otherwise 0. | 0.318 | 0.466 |
| *No Exemption* | It is 1 if student entered the school between 1889-1898 (period without exemption), otherwise, otherwise 0. | 0.479 | 0.499 |
| *Tokyo* | It is 1 if a student was born in Tokyo, otherwise 0. | 0.171 | 0.377 |
| *Commoner* | It is 1 if individual belonged to the commoner people in the Edo period, otherwise 0. | 0.725 | 0.446 |
| *Class size* [a] | Number of students in class where respondents belonged | 37.43 | 17.13 |
| *Name Change* | It is 1 if a student who changed his name, otherwise 0. | 0.035 | 0.183 |

Notes: Number of observations 39,694. Mean value of *Head* is simple average while that *Head Rate* is class size's weighted average.

**Table 3.** Analysis of academic performance comparing the Head and non-head (OLS):

|  | (1) Academic Performance | (2) Skip |
|---|---|---|
| Head ×Head_Exemption | 0.118** | 0.037** |
|  | (2.46) | (2.47) |
| Head ×No Exemption | 0.084* | 0.024* |
|  | (1.76) | (1.66) |
| Head_Exemption | 0.025*** | 0.012*** |
|  | (2.96) | (4.25) |
| No Exemption | 0.106*** | 0.020*** |
|  | (10.85) | (5.31) |
| Head | −0.091* | −0.018 |
|  | (−1.84) | (−1.15) |
| Name Change | 0.047*** | 0.016*** |
|  | (6.58) | (3.25) |
| $R^2$ | 0.04 | 0.16 |
| Observations | 39,694 | 39,694 |
| Number of groups | 1,672 | 1,672 |

Note: *, **, and *** indicate statistical significance at the 10%, 5%, and 1% levels, respectively. The numbers without parentheses are coefficients. Numbers in parentheses are t-values using the cluster-robust standard error. Number of cluster groups are equivalent to number of classes where students belonged. Number of groups are equivalent to number of classes. 1672 dummies for groups are included to control characteristics of classes where students belonged.

**Table 4.** Analysis of academic performance comparing the Head of Tokyo with Head of other regions (OLS):

|  | (1) *Academic Performance* | (2) *Skip* |
|---|---|---|
| *Head ×Head_Exemption ×Tokyo* | 0.243** (2.25) | 0.088 (0.80) |
| *Head ×No Exemption ×Tokyo* | 0.239** (2.23) | 0.112 (1.02) |
| *Head ×Head_Exemption* | 0.108** (2.22) | 0.036*** (2.65) |
| *Head ×No Exemption* | 0.077 (1.59) | 0.019 (1.55) |
| *Tokyo ×Head_Exemption* | 0.003 (0.28) | −0.003 (−0.66) |
| *Tokyo ×No Exemption* | 0.017 (1.64) | −0.003 (−0.81) |
| *Tokyo ×Head* | −0.267** (−2.53) | −0.117 (−1.07) |
| *Head_Exemption* | 0.019** (2.05) | 0.012*** (3.59) |
| *No Exemption* | 0.094*** (9.22) | 0.018*** (4.59) |
| *Tokyo* | −0.044*** (−5.22) | −0.009*** (−3.15) |
| *Head* | −0.082 (−1.64) | −0.013 (−0.98) |
| *Name Change* | 0.047*** (6.47) | 0.016*** (3.17) |
| $R^2$ | 0.05 | 0.17 |
| Observations | 39,694 | 39,694 |
| Number of groups | 1,672 | 1,672 |

Note: *, **, and *** indicate statistical significance at the 10%, 5%, and 1% levels, respectively. The numbers without parentheses are coefficients. Numbers in parentheses are t-values using the cluster-robust standard error. Number of cluster groups are equivalent to number of classes where students belonged. 1672 dummies for groups are included to control characteristics of classes where students belonged.

**Table 5.** Analysis of academic performance comparing the Head of Commoner with Head of Samurai (OLS):

|  | (1) Academic Performance | (2) Skip |
|---|---|---|
| Head ×Head_Exemption ×Commoner | 0.175 (1.45) | 0.084* (1.71) |
| Head ×No Exemption ×Commoner | 0.131 (1.08) | 0.067 (1.36) |
| Head ×Head_Exemption | 0.065 (1.27) | 0.014 (1.11) |
| Head ×No Exemption | 0.064 (1.23) | 0.014 (1.15) |
| Commoner ×Head_Exemption | −0.015* (−1.76) | −0.008* (−1.80) |
| Commoner ×No Exemption | 0.007 (0.86) | 0.001 (0.17) |
| Commoner ×Head | −0.158 (−1.32) | −0.080* (−1.68) |
| Head_Exemption | 0.035** (3.46) | 0.018*** (4.37) |
| No Exemption | 0.098*** (8.52) | 0.018*** (3.79) |
| Commoner | 0.007 (1.03) | 0.002 (0.63) |
| Head | −0.050 (−0.96) | 0.002 (0.20) |
| Name Change | 0.047*** (6.57) | 0.017*** (3.30) |
| $R^2$ | 0.04 | 0.16 |
| Observations | 39,694 | 39,694 |
| Number of groups | 1,672 | 1,672 |

Note: *, **, and *** indicate statistical significance at the 10%, 5%, and 1% levels, respectively. The numbers without parentheses are coefficients. Numbers in parentheses are t-values using the cluster-robust standard error. Number of cluster groups are equivalent to number of classes where students belonged. 1672 dummies for groups are included to control characteristics of classes where students belonged.

**Table 6.** Duration analysis for drop-out (Competing-risks regression):

|  | (1) | (2) | (3) | (4) | (5) | (6) |
|---|---|---|---|---|---|---|
|  |  | <=1897 |  |  | <=1898 |  |
| *Academic Performance* | 0.109*** |  |  | 0.136*** |  |  |
|  | (−14.9) |  |  | (−18.4) |  |  |
| *Skip* |  | 0.781* |  |  | 0.697** |  |
|  |  | (−1.76) |  |  | (−2.52) |  |
| *Head ×Head_Exemption* | 0.417*** | 0.360*** | 0.358*** | 0.412*** | 0.358*** | 0.355*** |
|  | (−2.74) | (−3.33) | (−3.35) | (−2.81) | (−3.36) | (−3.38) |
| *Head ×No Exemption* | 0.288*** | 0.263*** | 0.262*** | 0.301*** | 0.273*** | 0.272*** |
|  | (−4.03) | (−4.51) | (−4.51) | (−3.96) | (−4.43) | (−4.44) |
| *Head_Exemption* | 0.533*** | 0.535*** | 0.535*** | 0.519*** | 0.526*** | 0.525*** |
|  | (−8.36) | (−8.50) | (−8.53) | (−8.76) | (−8.67) | (−8.71) |
| *No Exemption* | 1.145* | 1.039 | 1.039 | 1.392*** | 1.287*** | 1.288*** |
|  | (1.66) | (0.47) | (0.48) | (3.87) | (2.88) | (2.88) |
| *Head* | 3.259*** | 3.301*** | 3.515*** | 3.266*** | 3.531*** | 3.534*** |
|  | (3.94) | (4.38) | (4.39) | (3.98) | (4.40) | (4.41) |
| *Class Size* | 1.015*** | 1.014*** | 1.014*** | 1.017*** | 1.015*** | 1.015*** |
|  | (7.39) | (6.55) | (6.55) | (7.50) | (6.79) | (6.78) |
| *Name Change* | 1.052 | 0.955 | 0.953 | 0.933 | 0.865* | 0.862* |
|  | (0.64) | (−0.57) | (−0.60) | (−0.84) | (−1.79) | (−1.83) |
| Wald-chi square | 572 | 241 | 240 | 938 | 223 | 222 |
| Observations | 38,867 | 38,867 | 38,867 | 39,694 | 39,694 | 39,694 |
| N. Failed (Drop-out) | 4,778 | 4,778 | 4,778 | 5,536 | 5,536 | 5,536 |
| N. competing (Graduate) | 1,095 | 1,095 | 1,095 | 1,164 | 1,164 | 1,164 |
| N. group (cluster) | 1,647 | 1,647 | 1,647 | 1,672 | 1,672 | 1,672 |

Note: **, and *** indicate statistical significance at the 5%, and 1% levels, respectively. The numbers without parentheses are sub-hazard ratios that can be interpreted similarly as hazard ratios in Cox regression. The numbers in parentheses are z-values using robust standard errors clustered on individuals. Number of cluster groups are equivalent to number of classes where students belonged.

# Appendix

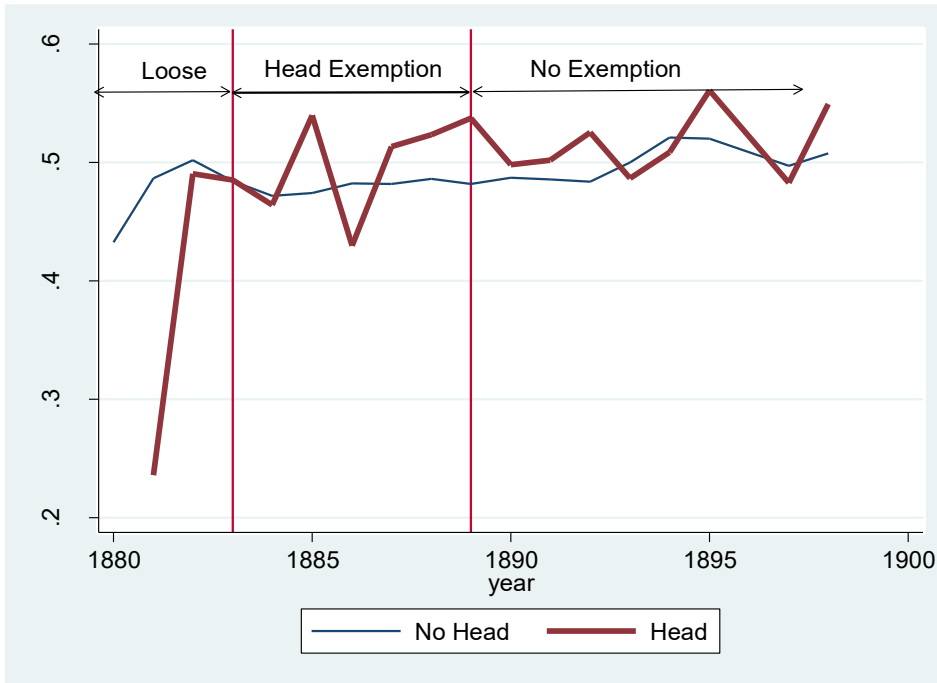

Figure A1. Trend check for Academic Performance: Head vs No-Head

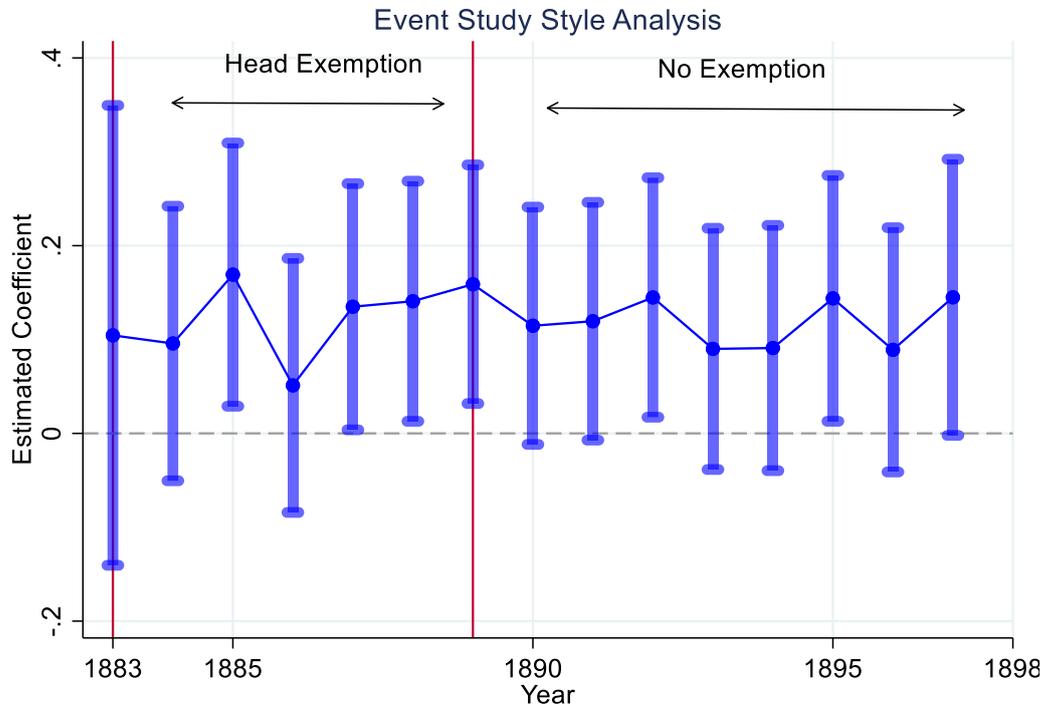

Figure A2. Coefficients obtained from Event-study specification.

Note: Group separated by years of entrance. Bars show 95% level confidence intervals.

Coefficients and Cis are results of specification as below;

$$Academic\ performance_{it} = \alpha_0 + \sum_{k=1883}^{1898} \alpha_k Head_i \times Years\ dummy_t + Z + e_{it}$$

Where Z is vector of year dummies and Head dummy. Reference group cover 1880-1882 period.

We reported coefficients of $\alpha_k$, interaction terms between Head and Year dummies.

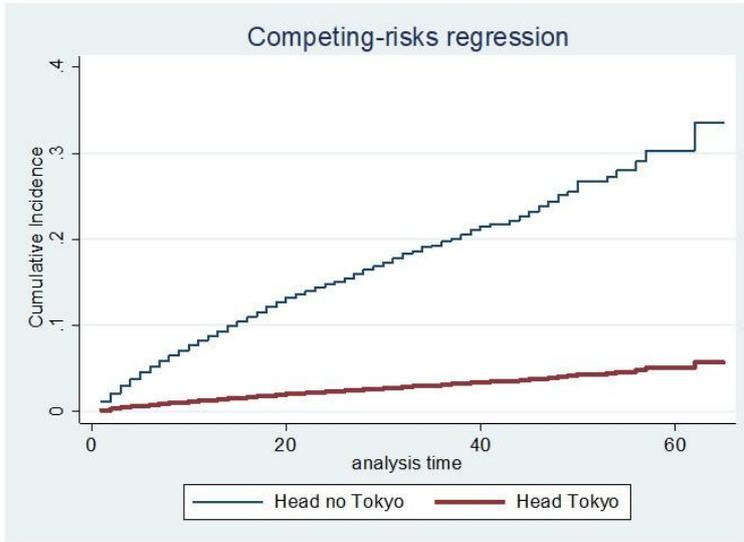

**FigA3. Drop-out Incidence Curve: Head of Tokyo vs Head of non_Tokyo 1884-1888 period**

(Drop-out and Graduation are as competing risks)

Note: Figure illustrated based on results of *Head×Head Exemption×Tokyo* and *Head×Head Exemption* and column (4) in Table A1.

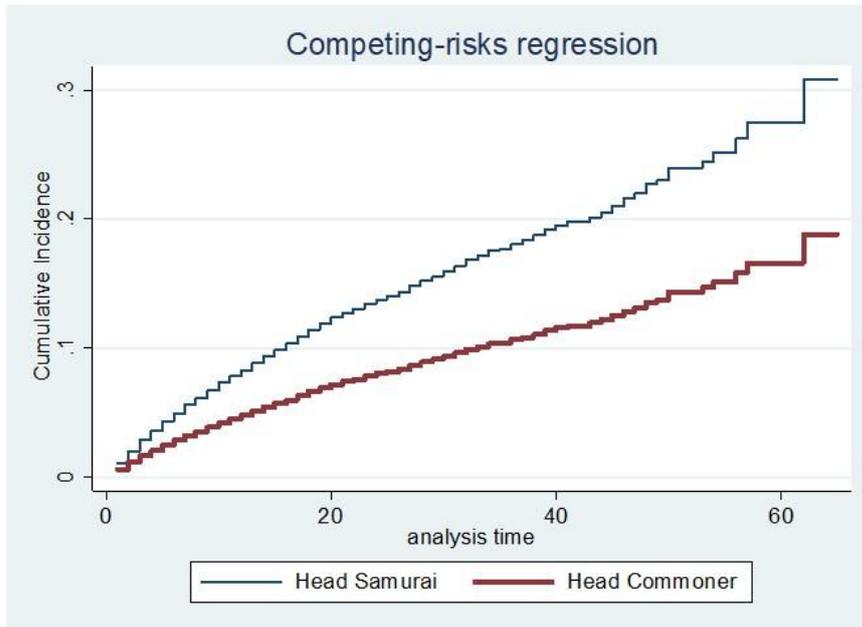

**Fig.A4. Drop-out Incidence Curve : Head of Commoner vs Head of Samurai**
(Drop-out and Graduation are as competing risks)
Note: Figure illustrated based on results of *Head ×Head Exemption ×Commoner* and *Head ×Head Exemption* and column (4) in Table A2.

**Table A1.** Duration analysis for drop-out comparing the Head of Tokyo with Head of other regions (Competing-risks regression):

|  | (1) | (2) | (3) | (4) | (5) | (6) |
|---|---|---|---|---|---|---|
|  |  | <=1897 |  |  | <=1898 |  |
| *Academic Performance* | 0.111*** | | | 0.137*** | | |
|  | (−13.81) | | | (−17.29) | | |
| *Skip* | | 0.756** | | | 0.680*** | |
|  | | (−1.98) | | | (−2.70) | |
| *Head ×Head_Exemption ×Tokyo* | 0.084*** | 0.079*** | 0.058*** | 0.078*** | 0.056*** | 0.056*** |
|  | (−5.37) | (−5.60) | (−5.59) | (−5.44) | (−5.61) | (−5.59) |
| *Head ×No Exemption ×Tokyo* | 0.079*** | 0.054*** | 0.055*** | 0.089*** | 0.062*** | 0.062*** |
|  | (−5.81) | (−5.97) | (−5.97) | (−5.75) | (−5.89) | (−5.88) |
| *Head ×Head_Exemption* | 0.586 | 0.509** | 0.504** | 0.581 | 0.507** | 0.502** |
|  | (−1.52) | (−2.00) | (−2.03) | (−1.55) | (−2.02) | (−2.05) |
| *Head ×No Exemption* | 0.412*** | 0.376*** | 0.374*** | 0.420** | 0.383*** | 0.380*** |
|  | (−2.60) | (−2.99) | (−3.01) | (−2.58) | (−2.98) | (−2.99) |
| *Tokyo ×Head_Exemption* | 1.878*** | 1.850*** | 1.851*** | 1.943*** | 1.893*** | 1.895*** |
|  | (5.36) | (5.26) | (5.27) | (5.71) | (5.47) | (5.48) |
| *Tokyo ×No Exemption* | 2.284*** | 2.060*** | 2.058*** | 2.543*** | 2.299*** | 2.297*** |
|  | (7.75) | (6.94) | (6.94) | (8.86) | (7.90) | (7.89) |
| *Tokyo ×Head* | 10.93*** | 15.78*** | 15.81*** | 11.46*** | 15.85*** | 15.89*** |
|  | (6.24) | (6.35) | (6.35) | (6.25) | (6.28) | (6.28) |
| *Head_Exemption* | 0.468*** | 0.475*** | 0.474*** | 0.451*** | 0.462*** | 0.461*** |
|  | (−9.48) | (−9.62) | (−9.66) | (−10.01) | (−9.86) | (−9.91) |
| *No Exemption* | 0.969 | 0.905 | 0.906 | 1.149 | 1.092 | 1.093 |
|  | (−0.37) | (−1.19) | (−1.18) | (1.61) | (1.00) | (1.02) |
| *Tokyo* | 0.397*** | 0.414*** | 0.415*** | 0.399*** | 0.413*** | 0.414*** |
|  | (−10.85) | (−10.82) | (−10.80) | (−11.09) | (−10.81) | (−10.79) |
| *Head* | 2.344** | 2.515*** | 2.527*** | 2.336** | 2.518*** | 2.533** |
|  | (2.57) | (2.92) | (2.93) | (2.58) | (2.93) | (2.05) |
| *Class Size* | 1.016*** | 1.013*** | 1.013*** | 1.017*** | 1.015*** | 1.015*** |
|  | (7.53) | (6.56) | (6.55) | (7.68) | (6.84) | (6.83) |
| *Name Change* | 1.034 | 0.944 | 0.941 | 0.914 | 0.854* | 0.851** |
|  | (0.38) | (−0.72) | (−0.75) | (−1.08) | (−1.95) | (−1.99) |
| Wald-chi square | 1,128 | 509 | 509 | 1,069 | 480 | 541 |
| Observations | 38,867 | 38,867 | 38,867 | 39,694 | 39,694 | 39,694 |
| N. Failed (Drop-out) | 4,778 | 4,778 | 4,778 | 5,536 | 5,536 | 5,536 |
| N. competing (Graduate) | 1,095 | 1,095 | 1,095 | 1,164 | 1,164 | 1,164 |
| N. group (cluster) | 1,647 | 1,647 | 1,647 | 1,672 | 1,672 | 1,672 |

Note: **, and *** indicate statistical significance at the 5%, and 1% levels, respectively. The numbers without parentheses are sub-hazard ratios that can be interpreted similarly as hazard ratios in Cox regression. The numbers in parentheses are z-values using robust standard errors clustered on individuals. Number of cluster groups are equivalent to number of classes where students belonged.

**Table A2.** Duration analysis for drop-out comparing the Head of Commoner with Head of Samurai (Competing-risks regression):

| | (1) | (2) | (3) | (4) | (5) | (6) |
|---|---|---|---|---|---|---|
| | | <=1897 | | | <=1898 | |
| *Academic Performance* | 0.109*** | | | 0.137*** | | |
| | (−14.83) | | | (−18.39) | | |
| *Skip* | | 0.779* | | | 0.696** | |
| | | (−1.78) | | | (−2.53) | |
| *Head ×Head_Exemption ×Commoner* | 0.313** | 0.190*** | 0.189*** | 0.303** | 0.192*** | 0.191*** |
| | (−2.03) | (−3.29) | (−3.30) | (−2.13) | (−3.28) | (−3.29) |
| *Head ×No Exemption ×Commoner* | 0.239** | 0.164*** | 0.137*** | 0.254** | 0.178*** | 0.178*** |
| | (−2.53) | (−3.64) | (−3.64) | (−2.47) | (−3.50) | (−3.51) |
| *Head ×Head_Exemption* | 0.555 | 0.536 | 0.533 | 0.550 | 0.530 | 0.526 |
| | (−1.28) | (−1.44) | (−1.46) | (−1.31) | (−1.48) | (−1.49) |
| *Head ×No Exemption* | 0.469* | 0.440** | 0.438** | 0.459* | 0.430** | 0.427** |
| | (−1.65) | (−1.91) | (−1.91) | (−1.72) | (−1.98) | (−1.99) |
| *Commoner ×Head_Exemption* | 0.739*** | 0.765*** | 0.766*** | 0.733*** | 0.762*** | 0.763*** |
| | (−3.45) | (−2.95) | (−2.94) | (−3.46) | (−2.98) | (−2.97) |
| *Commoner ×No Exemption* | 0.829** | 0.836** | 0.837** | 0.918 | 0.930 | 0.931 |
| | (−2.16) | (−2.06) | (−2.05) | (−0.99) | (−0.84) | (−0.83) |
| *Commoner ×Head* | 3.515** | 5.416*** | 5.442*** | 3.648** | 5.429*** | 5.464*** |
| | (2.33) | (3.63) | (3.64) | (2.44) | (3.64) | (3.65) |
| *Head_Exemption* | 0.645*** | 0.630*** | 0.629*** | 0.629*** | 0.620*** | 0.618*** |
| | (−4.73) | (−5.01) | (−5.03) | (−5.01) | (−5.16) | (−5.19) |
| *No Exemption* | 1.266*** | 1.143 | 1.143 | 1.413*** | 1.296** | 1.296** |
| | (2.37) | (1.34) | (1.34) | (3.53) | (2.61) | (2.61) |
| *Commoner* | 1.267*** | 1.259*** | 1.258*** | 1.266*** | 1.255*** | 1.253*** |
| | (3.89) | (3.79) | (3.77) | (3.86) | (3.72) | (3.70) |
| *Head* | 2.295* | 2.300** | 2.307** | 2.283* | 2.305** | 2.314** |
| | (1.91) | (2.05) | (2.06) | (1.91) | (2.07) | (2.08) |
| *Class Size* | 1.015*** | 1.013*** | 1.013*** | 1.017*** | 1.015*** | 1.015*** |
| | (7.29) | (6.45) | (6.44) | (7.39) | (6.69) | (6.68) |
| *Name Change* | 1.045 | 0.946 | 0.944 | 0.922 | 0.854* | 0.852** |
| | (0.55) | (−0.68) | (−0.71) | (−0.97) | (−1.94) | (−1.98) |
| Wald-chi square | 739 | 612 | 613 | 1,260 | 541 | 479 |
| Observations | 38,867 | 38,867 | 38,867 | 39,694 | 39,694 | 39,694 |
| N. Failed (Drop-out) | 4,778 | 4,778 | 4,778 | 5,536 | 5,536 | 5,536 |
| N. competing (Graduate) | 1,095 | 1,095 | 1,095 | 1,164 | 1,164 | 1,164 |
| N. group (cluster) | 1,647 | 1,647 | 1,647 | 1,672 | 1,672 | 1,672 |

Note: *, **, and *** indicate statistical significance at the 10%, 5%, and 1% levels, respectively. The numbers without parentheses are sub-hazard ratios that can be interpreted similarly as hazard ratios in Cox regression. The numbers in parentheses are z-values using robust standard errors clustered on individuals. Number of cluster groups are equivalent to number of classes where students belonged.